\documentclass[%
reprint,
superscriptaddress,
nofootinbib,
amsmath,
amssymb,
aps,
floatfix,
showkeys,
]{revtex4-2}

\usepackage[colorlinks,citecolor=blue,linkcolor=blue,anchorcolor=blue,filecolor=blue,
urlcolor=blue]{hyperref}

\usepackage{graphicx,color,xcolor}
\usepackage{amsfonts,amsmath}
\usepackage{amssymb,ulem}
\usepackage{dcolumn}
\usepackage{bm,lipsum}
\usepackage[utf8]{inputenc}
\usepackage{subeqnarray}
\usepackage{array}
\usepackage{epsfig,hyperref}
\usepackage{morefloats}
\usepackage{relsize}
\usepackage{url}

\usepackage{comment}

\usepackage{cleveref}
\usepackage{orcidlink}

\begin{document}

\preprint{ }

\title{Baryon and Pseudoscalar Meson Octets within a Unified broken SU(6) Symmetry}

\author{Luiz L. Lopes}~%\orcidlink{0000-0003-0982-9774}}
\email{llopes@cefetmg.br}

\affiliation{ Centro Federal de Educac\~{a}o Tecnol\'{o}gica de Minas Gerais Campus VIII; \\ CEP 37.022-560, Varginha - MG - Brazil\\}
% \mbox{$^2$Depto de F\'{\i}sica - CFM - Universidade Federal de Santa Catarina  Florian\'opolis - SC - CP. 476 - CEP 88.040 - 900 - Brazil }}

\date{\today}

\begin{abstract}
In this work, I discuss neutron stars with hyperons and anti-kaon condensate. To fix their coupling constants with the vector mesons of the Quantum Hadrodynamics, I use a unified scheme imposing that the Yukawa coupling is an invariant under SU(3) and SU(6) groups.  Combining with the G-Parity, I show that some expected results of the kaon and anti-kaon interaction with the nucleus are re-obtained. In the same sense, the naive quark-isospin counting rule is restored in the SU(6) limit.
Furthermore, the G-Parity combined with the SU(3) gives us a clear picture of the role played by each meson in the kaon condensation. Numerical results show that the presence of anti-kaons severely compromises the stiffening of the EOS by breaking the SU(6) symmetry.

%Keywords: Dark Matter, Neutron Star, Compact Objects %\sep Física de partículas \sep 

\end{abstract}

\maketitle

\section{Introduction}

Proposed even before the existence of the neutron by Lev. Landau~\cite{Landau1932}, neutron stars are exotic objects, where all physical quantities, such as density, gravity, and magnetic field, can reach values far greater than we can obtain in terrestrial laboratories.

Due to the high densities reached in neutron stars' interiors, their internal composition is still a mystery. The possibility of hyperons~\cite{GLENDENNING1982,WeissPRC2012,Miyatsu2013,lopesnpa,lopes2023ptep}, $\Delta$'s resonances~\cite{BETHE1974,Kauan2022PRC,lopesPRD}, kaon condensate~\cite{kaonp,Banik2001,Thapa2021,Ma2022,Thakur2025}, and even deconfined quark-gluon plasma~\cite{Maruyama2007} in the neutron stars' core were already studied in the literature. The biggest problem when new degrees of freedom are introduced is that our knowledge about their interaction with nuclear matter is very low. In the case of hyperons, the potential depths are known with satisfactory accuracy. But yet, different parameters can produce the same potential depth, making the result strongly model-dependent. In the case of kaons and $\Delta$'s resonances, the situation is even worse, once the uncertainties about the true value of their potential depths are large.

One way to overcome such difficulties is via symmetry group arguments by imposing that the Yukawa Lagrangian~\cite{Yukawa1935} is an invariant under SU(3) flavor symmetry. This technique was successfully used in the past to fix the vector meson coupling constant with the baryon octet~\cite{WeissPRC2012,Miyatsu2013,lopesnpa,lopes2023ptep} and the baryon decuplet~\cite{lopesPRD}. With this approach, all baryon-vector meson coupling constants can be determined in terms of three free parameters $z$, $\theta_V$, and $\alpha_V$~\cite{DOVER1984}.

In this work, I use the same technique to fix the coupling constants of the vector mesons with the pseudo-scalar octet. Unlike the baryon octet, where each member is a different baryon,  there are particles and antiparticles in the pseudo-scalar octet. Due to this fact, I can use the G-Parity to help constrain the free parameters related to the SU(3) flavor symmetry~\footnote{A good discussion about the concept and applications of the G-Parity is given in Section 5.6 of ref.~\cite{GreinerSymmetries}.}. I show that combining SU(3) symmetry with G-Parity, some experimental results ( as the kaons feel a repulsive potential while anti-kaons feel an attractive one) are recovered. Moreover, the naive quark-isospin counting rule is also restored in the SU(6) limit.

The paper is organized as follows: in Section II, I present the formalism of Quantum Hadrodynamics (QHD) for baryons and mesons in the mean-field approximation (MFA), calculating their energy eigenvalues, energy densities, and the mesonic expectation values. In Section III, I present the coupling constants calculated within the SU(3) formalism for the baryon octet, while in Section  IV the I present the results for the pseudoscalar meson octet. I also apply the G-Parity to remove all the free parameters of the SU(3) formalism. The numerical results are presented in Section V.

\section{Formalism}\label{fm}

I start with a modified version of the  QHD Lagrangian~\cite{Serot_1992}, with the $\sigma\omega\rho\phi$ mesons as well as non-linear terms. It  reads~\cite{Miyatsu2013,lopesPRD,LopesUNIVERSE2025}:

\begin{eqnarray}
\mathcal{L} =  \sum_B\bar{\psi}_B[\gamma^\mu(\mbox{i}\partial_\mu  - g_{BB\omega}\omega_\mu  -g_{BB\phi}\phi_\mu   - g_{BB\rho} \frac{1}{2}\vec{\tau} \cdot \vec{\rho}_\mu) \nonumber \\
- (M_B - g_{BB\sigma}\sigma)]\psi_B  -U(\sigma) + \frac{1}{2}(\partial_\mu \sigma \partial^\mu \sigma - m_s^2\sigma^2)  \nonumber   \\
   - \frac{1}{4}\Omega^{\mu \nu}\Omega_{\mu \nu} + \frac{1}{2} m_v^2 \omega_\mu \omega^\mu+ \Lambda_{\omega\rho}(g_{NN\rho}^2 \vec{\rho^\mu} \cdot \vec{\rho_\mu}) (g_{NN\omega}^2 \omega^\mu \omega_\mu)  \nonumber \\
- \frac{1}{4}\Phi^{\mu \nu}\Phi_{\mu \nu} + \frac{1}{2} m_\phi^2 \phi_\mu \phi^\mu + \frac{1}{2} m_\rho^2 \vec{\rho}_\mu \cdot \vec{\rho}^{ \; \mu} - \frac{1}{4}\bf{P}^{\mu \nu} \cdot \bf{P}_{\mu \nu}  , \nonumber \\ \label{QHD} 
\end{eqnarray}
in natural units. The $\psi_B$ are the Dirac field of the baryon $B$ of mass $M_B$. The sum in $B$ can run only for the nucleons or over the entire baryon octet. The g's are the coupling constants between the baryons and mesons. The $i = \sigma,\omega,\rho,\phi$ are the mesonic field with mass $m_i$, and $\bar{\tau}$ are the Pauli matrices. The antisymmetric mesonic field strength tensors are given by their usual expressions~\cite{Glenbook}. The $\Lambda_{\omega\rho}$ determines the strength of the coupling between the $\omega$ and $\rho$ mesons as introduced in ref.~\cite{IUFSU}. Finally, $U(\sigma)$ is the non-linear term introduced in ref.~\cite{Boguta} to correct the incompressibility:

\begin{equation}
  U(\sigma) = \frac{\kappa M_N(g_{NN\sigma}\sigma)^3}{3} + \frac{\lambda(g_{NN\sigma}\sigma)^4}{4}  .
\end{equation}

Applying the Euler-Lagrange equation and solving the Dirac equation, the energy eigenvalue is obtained. In MFA, we have:

\begin{equation}
E_B =  \sqrt{M^{*2}_B + k^2} + g_{BB\omega}\omega_0 + g_{BB\phi}\phi_0 + \frac{\tau_3}{2} g_{BB\rho}\rho_0 \label{eigen} ,
\end{equation}
where $M^*_B = M_B - g_{BB\sigma}\sigma_0$ is the effective mass of the baryon B.
At $ T=0$ K, the energy eigenvalue is also the chemical potential of the baryon $B$, $\mu_B$.

Another important quantity is the potential depth of the baryon B, defined at the saturation density~\cite{Potentials2000}:

\begin{equation}
    U_B(n_0) = g_{BB\omega}\omega_0 - g_{BB\sigma}\sigma_0.
\end{equation}

As neutron stars are stable macroscopic objects, we need to describe a neutral, chemically stable matter, and hence, leptons are added as free Fermi gases. 
The energy density for baryons and leptons is given as:

 \begin{eqnarray}
  \epsilon = \frac{1}{\pi^2} \sum_B\int_0^{k_{fB}} [\sqrt{M_B^{*2} +k^2}]  k^2 dk \nonumber  + \frac{1}{2}m_s^2\sigma_0^2 \nonumber \\ + \frac{1}{2}m_\omega^2\omega_0^2  + \frac{1}{2}m_\rho^2\rho_0^2  \nonumber  +  \frac{\kappa M_N (g_{NN\sigma}\sigma_0)^3}{3} + \frac{\lambda (g_{NN\sigma}\sigma_0)^4}{4} \\+3\Lambda_v\omega_0^2\rho_0^2  + \frac{1}{\pi^2}\sum_l\int_0^{k_{fl}}\sqrt{m_l^{2} +k^2} k^2 dk  \nonumber , \\ \label{iufsued2}
 \end{eqnarray}
 where $\Lambda_v = \Lambda_{\omega\rho}g_{NN\omega}^2g_{NN\rho}^2$. Furthermore, the sum in $B$ runs over the baryons and the sum in $l$ over the leptons.

 The Lagrangian for the members of the pseudoscalar-meson octet is~\cite{kaonp,Banik2001,Thapa2021,Ma2022,Thakur2025}:

\begin{equation}
\mathcal{L}_M = D_\mu^{*}\bar{M}D^\mu M - m_M^{*}\bar{M}M ,
\end{equation}
where $M$ is the mesonic field, $D_\mu = \partial_\mu + ig_{MM\omega}\omega_\mu + \frac{i}{2}g_{MM\rho}\vec{\tau} \cdot \vec{\rho}_\mu + i g_{MM\phi}\phi_\mu$ the covariant derivative, and $m_M^{*} = m_M - g_{MM\sigma}\sigma$ is the effective mass of the meson $M$. By applying the Euler-Lagrange equations, the energy eigenvalue, also known as the dispersion relation~\cite{Banik2001}, is obtained. In MFA, for the S-wave condensate (zero momentum), we have: 

\begin{equation}
  \omega_M = m^*_M + g_{MM\omega}\omega_0 + \frac{\tau_{3}}{2}g_{MM\rho}\rho_0 + g_{MM\phi}\phi_0  . \label{disprel}
\end{equation}

In the same sense, the potential depth of the meson M at the saturation density is:

\begin{equation}
U_M(n_0) =  g_{MM\omega}\omega_0 - g_{MM\sigma}\sigma_0. \label{mesonpd}
\end{equation}

The expected values of the virtual mesonic fields in MFA are given by:

\begin{eqnarray}
 m_s^2\sigma_0 + g_{NN\sigma}[\kappa M_N (g_{NN\sigma}\sigma_0)^2  +\lambda(g_{NN\sigma}\sigma_0)^3] \nonumber \\   = \sum_B g_{BB\sigma}n^S_B + \sum_M g_{MM\sigma}n_M ,
\end{eqnarray}
\begin{equation}
 (m_\omega^2 + 2\Lambda_v\rho_0^2)\omega_0 = \sum_B g_{BB\omega}n_B + \sum_M g_{MM\omega}n_M  ,
\end{equation}

\begin{equation}
 (m_\rho^2 + 2\Lambda_v\omega_0^2)\rho_0 = \sum_B g_{BB\rho}\frac{\tau_3}{2}n_B + \sum_M g_{MM\rho}\frac{\tau_3}{2}n_M  
\end{equation}

\begin{equation}
  m_\phi^2\phi_0 = \sum_B g_{BB\phi}n_B + \sum_M g_{MM\phi}n_M ,
\end{equation}
where $n_B^S,~n_B$, and $n_M$ are, respectively, the scalar density, the number density, and the meson density, given by~\cite{kaonp}:

\begin{equation}
n^S_B = \frac{1}{\pi}\int_0^{k_{fB}} \frac{M^*_B}{\sqrt{M_B^{*2} + k^2}} k^2 dk ,
\end{equation}

\begin{equation}
   n_B = \frac{k_{fB}^2}{3\pi^2} ,
\end{equation}

\begin{equation}
n_M = 2m_M^*\bar{M}M.   
\end{equation}

 The mesonic energy density is

 \begin{equation}
    \epsilon_M =  \sum_M m^{*}_Mn_M ,
 \end{equation}

The total energy density is the sum of the baryonic, leptonic, and mesonic energy densities. The pressure is obtained via thermodynamic relations: $p = \sum_f\mu_fn_f - \epsilon_T$, where the sum runs over all the fermions.

Now, considering a chemically stable matter with zero net charge, composed of nucleons, hyperons, leptons, and anti-kaon condensate, we have:

\begin{equation}
 \mu_B = \mu_n - q_B\mu_e; \quad \mu_e =\mu_\mu = \mu_{K^-} = \omega_{K^-};    \nonumber
\end{equation}

\begin{equation}
    \omega_{\bar{K^0}} = 0; \quad \sum_Bq_Bn_B -\sum_ln_l -n_{k^-} = 0,
\end{equation}
where $q_B = \pm 1$ is the electric charge of the baryon $B$ in terms of the proton charge.

\section{SU(3) and SU(6) symmetry groups for the hyperons-vector mesons coupling constants}\label{pr}

I assume that the Yukawa Lagrangian of the QHD is invariant under the SU(3) flavor symmetry:

\begin{equation}
  \mathcal{L}_{YUK} = -g(\bar{\psi}_B\psi_B)M , \label{yuk}
\end{equation}
in other words, Eq.~\ref{yuk} is a unitary singlet. 

Therefore, the coupling constants for different baryons and mesons are not independent of each other. The calculation of different coupling constants involves the calculation of the SU(3) Clebsch-Gordan (CG) coefficients.
The baryon octet is divided into a nucleon doublet, N =$\{$p,n$\}$, a $\Sigma$ triplet $\Sigma =\{\Sigma^-,\Sigma^0,\Sigma^+\}$, a $\Lambda$ singlet, $\Lambda =\{\Lambda^0\}$, and a $\Xi$ doublet, $\Xi = \{\Xi^-,\Xi^0\}$.
Their representation in the hypercharge $(Y)$ isospin projection $I_3$ scheme is presented below.

%\begin{figure}[t]
%\centering
%\includegraphics[width=0.333\textwidth,angle=0]{BO.eps}
%\caption{Baryon octet in the hypercharge ($Y$) isospin projection $I_3$ %scheme.}\label{BO}
%\end{figure}

\begin{center}
\begin{tikzpicture}[scale=2, thick]

% Eixos
\draw[->, gray] (-1.5, 0) -- (1.5, 0) node[right] {$I_3$};
\draw[->, gray] (0, -1.5) -- (0, 1.5) node[above] {$Y$};

% Pontos
\fill (0.5,1) circle (2.2pt);    % p
\fill (-0.5,1) circle (2.2pt);   % n
\fill (1,0.0) circle (2.2pt);    % Σ⁺ (levemente acima)
\fill (-1,0.0) circle (2.2pt);   % Σ⁻ (levemente acima)
\fill (0.5,-1) circle (2.2pt);   % Ξ⁰
\fill (-0.5,-1) circle (2.2pt);  % Ξ⁻
\fill (0,0) circle (2.2pt);      % Σ⁰ e Λ⁰ no mesmo ponto

% Conexões corretas
\draw[gray] (0.5,1) -- (-0.5,1);          % p <-> n
\draw[gray] (0.5,1) -- (1,0.0) -- (0.5,-1);  % p -> Σ⁺ -> Ξ⁰
\draw[gray] (-0.5,1) -- (-1,0.0) -- (-0.5,-1); % n -> Σ⁻ -> Ξ⁻
%\draw[gray] (-1,0.1) -- (1,0.1);         % Σ⁻ <-> Σ⁺ (acima do centro)
\draw[gray] (-0.5,-1) -- (0.5,-1);       % Ξ⁻ <-> Ξ⁰

% Rótulos
\node at (0.65,1.1) {$p$};
\node at (-0.65,1.1) {$n$};
\node at (1.3,0.2) {$\Sigma^+$};
\node at (-1.3,0.2) {$\Sigma^-$};
\node at (0.65,-1.1) {$\Xi^0$};
\node at (-0.65,-1.1) {$\Xi^-$};

% Rótulos no centro
\node at (-0.35,0.25) {$\Sigma^0$};
\node at (0.4,-0.25) {$\Lambda^0$};

% Título
\node at (0,1.8) {\textbf{Baryon Octet (SU(3))}};

\end{tikzpicture}
\end{center}

The techniques to calculate the CG for the baryon octet and the members of the vector meson octet are well-known and are presented in detail in ref.~\cite{Swart1963}. The use of the SU(3) CG coefficients to describe hyperonic neutron stars is also well-known in the literature. It is detailed in the Appendix of ref.~\cite{lopesPRD} and in ref.~\cite{lopes2023ptep}:

\begin{equation}
  g_{NN\rho} = g_8,  
\end{equation}

\begin{equation}
  g_{\Sigma\Sigma\rho} = 2g_8\alpha_V,  
\end{equation}

\begin{equation}
  g_{\Xi\Xi\rho} =- g_8(1 - 2\alpha_V),  
\end{equation}

\begin{equation}
  g_{\Lambda\Lambda\rho} = 0,  
\end{equation}

\begin{equation}
  g_{\Sigma\Lambda\rho} = \frac{2}{3}\sqrt{3}g_8(1 -\alpha_V),  
\end{equation}

\begin{equation}
  g_{NN\omega} = g_1\cos\theta_V + g_8\sin\theta_V \frac{1}{3}\sqrt{3}(4\alpha_V -1),  
\end{equation}

\begin{equation}
  g_{\Lambda\Lambda\omega} = g_1\cos\theta_V - g_8\sin\theta_V \frac{2}{3}\sqrt{3}(1 - \alpha_V),  
\end{equation}

\begin{equation}
  g_{\Sigma\Sigma\omega} = g_1\cos\theta_V + g_8\sin\theta_V \frac{2}{3}\sqrt{3}(1 - \alpha_V),  
\end{equation}

\begin{equation}
  g_{\Xi\Xi\omega} = g_1\cos\theta_V - g_8\sin\theta_V \frac{1}{3}\sqrt{3}(1 + 2\alpha_V). 
\end{equation}

The results for the $\phi$ meson is obtained by replacing $\sin\theta_V \to \cos\theta_V$ and $\cos\theta_V \to  -\sin\theta_V$ in the equations related to the $\omega$ meson~\cite{DOVER1984}.

The SU(3) symmetry let us with three free parameters, the $\theta_V$ related to the nature of the $\omega$ and $\phi$ mesons; $z = g_8/g_1$, related to the relative strength of the octet and the singlet states, and $\alpha_V = F/(F+D)$, related to the relative strength of the antisymmetric coupling $(F)$ in relation to the symmetric one ($D$). Additional discussion can be found in refs.~\cite{lopes2023ptep,lopesPRD,DOVER1984,Swart1963} and the references therein.

These three free parameters can be fixed by imposing that the Yukawa Lagrangian is not only invariant under the SU(3) flavor symmetry, but also the SU(2) spin symmetry, raising the hybrid SU(6)$\supset$SU(3)$\otimes$SU(2) group. In the SU(6) limit we have:

\begin{equation}
  z =\frac{1}{\sqrt{6}}, \quad   \theta_V = 35.264º, \quad \alpha_V = 1.00,
\end{equation}
which implies $\phi = \langle s\bar{s}\rangle$, $g_{NN\phi} = 0$, and the restoring of the naive quark-isospin counting rule.

Here, I assume that the SU(3) symmetry is kept but the more restrictive SU(6) symmetry is partially broken, i.e, $z$ and $\theta_V$ assume their values in accordance with the SU(6) symmetry, but $\alpha_V$ is left as a free parameter.
The relative strength of the coupling constants of the hyperons in relation to the nucleon for the vector mesons are:

\begin{eqnarray}
  \frac{g_{\Sigma\Sigma\rho}}{g_{NN\rho}} = 2\alpha_V; \quad \frac{g_{\Xi\Xi\rho}}{g_{NN\rho}} =   (2\alpha_V - 1); \nonumber \\
   \frac{g_{\Lambda\Lambda\rho}}{g_{NN\rho}} = 0.0; \quad  \frac{g_{\Sigma\Lambda\rho}}{g_{NN\rho}} = \frac{2}{3}\sqrt{3}(1 - \alpha_V) ;  \end{eqnarray}

\begin{eqnarray}
  \frac{g_{\Sigma\Sigma\omega}}{g_{NN\omega}} =  \frac{8 -2\alpha_V}{5 + 4\alpha_V}; \quad \frac{g_{\Xi\Xi\omega}}{g_{NN\omega}} =   \frac{5 -2\alpha_V}{5 + 4\alpha_V}; \nonumber \\
   \frac{g_{\Lambda\Lambda\omega}}{g_{NN\omega}} = \frac{4 + 2\alpha_V}{5 + 4\alpha_V};;  \end{eqnarray}

\begin{eqnarray}
  \frac{g_{\Sigma\Sigma\phi}}{g_{NN\omega}} =  -\sqrt{2} \bigg (\frac{8 -2\alpha_V}{5 + 4\alpha_V} \bigg );  \quad  \frac{g_{\Xi\Xi\phi}}{g_{NN\omega}} =   -\sqrt{2} \bigg (\frac{8 -2\alpha_V}{5 + 4\alpha_V} \bigg ); \nonumber \\
   \frac{g_{\Lambda\Lambda\phi}}{g_{NN\omega}} = -\sqrt{2} \bigg (\frac{8 -2\alpha_V}{5 + 4\alpha_V} \bigg );  \quad  \frac{g_{NN\phi}}{g_{NN\omega}} = 0.0; \nonumber \\ 
   \end{eqnarray}

\section{SU(3) and SU(6) symmetry groups for the pseudoscalar-vector meson couplings}\label{pm}

Analogous to the baryon octet, I assume that the Yukawa Lagrangian of the QHD for the members of the pseudoscalar octet is also invariant under the SU(3) flavor symmetry:

\begin{equation}
  \mathcal{L}_{YUK} = -g(\bar{M}_{PS}M_{PS})M_V , \label{yuk2}
\end{equation}
where the subscripts $PS$ and $V$ point to whether the mesons are members of the pseudoscalar or vector octets.

The members of the pseudoscalar octet are analogous to the baryon octet, and can be divided into a kaon doublet, K =$\{K^0,K^+\}$, a $\pi$ triplet $\pi =\{\pi^-,\pi^0,\pi^+\}$, a $\eta$ singlet, $\eta =\{\eta_8\}$, and a anti-kaon doublet, $\bar{K} = \{K^-,\bar{K}^0\}$.
Their representation in the hypercharge $(Y)$ isospin projection $I_3$ scheme is exactly the same as the baryon octet, as can be seen below.

%\begin{figure}[t]
%\centering
%\includegraphics[width=0.333\textwidth,angle=0]{BO.eps}
%\caption{Baryon octet in the hypercharge ($Y$) isospin projection $I_3$ %scheme.}\label{BO}
%\end{figure}

\begin{center}
\begin{tikzpicture}[scale=2, thick]

% Eixos
\draw[->, gray] (-1.5, 0) -- (1.5, 0) node[right] {$I_3$};
\draw[->, gray] (0, -1.5) -- (0, 1.5) node[above] {$Y$};

% Pontos
\fill (0.5,1) circle (2.2pt);    % p
\fill (-0.5,1) circle (2.2pt);   % n
\fill (1,0.0) circle (2.2pt);    % Σ⁺ (levemente acima)
\fill (-1,0.0) circle (2.2pt);   % Σ⁻ (levemente acima)
\fill (0.5,-1) circle (2.2pt);   % Ξ⁰
\fill (-0.5,-1) circle (2.2pt);  % Ξ⁻
\fill (0,0) circle (2.2pt);      % Σ⁰ e Λ⁰ no mesmo ponto

% Conexões corretas
\draw[gray] (0.5,1) -- (-0.5,1);          % p <-> n
\draw[gray] (0.5,1) -- (1,0.0) -- (0.5,-1);  % p -> Σ⁺ -> Ξ⁰
\draw[gray] (-0.5,1) -- (-1,0.0) -- (-0.5,-1); % n -> Σ⁻ -> Ξ⁻
%\draw[gray] (-1,0.1) -- (1,0.1);         % Σ⁻ <-> Σ⁺ (acima do centro)
\draw[gray] (-0.5,-1) -- (0.5,-1);       % Ξ⁻ <-> Ξ⁰

% Rótulos
\node at (0.70,1.15) {$K^+$};
\node at (-0.70,1.15) {$K^0$};
\node at (1.3,0.2) {$\pi^+$};
\node at (-1.3,0.2) {$\pi^-$};
\node at (0.65,-1.1) {$\bar{K^0}$};
\node at (-0.65,-1.1) {$K^-$};

% Rótulos no centro
\node at (-0.35,0.25) {$\pi^0$};
\node at (0.4,-0.25) {$\eta_8$};

% Título
\node at (0,1.8) {\textbf{Pseudoscalar Octet (SU(3))}};

\end{tikzpicture}
\end{center}

As the quantum numbers of the baryon octet are the same as the pseudoscalar octet, their CG coefficients are also the same:

\begin{equation}
  g_{KK\rho} = g_{8(PS)},   \label{krho}
\end{equation}

\begin{equation}
  g_{\bar{K}\bar{K}\rho} =- g_{8(PS)}(1 - 2\alpha_{V(PS)}),  \label{kbarrho}
\end{equation}

\begin{equation}
  g_{\pi\pi\rho} = 2g_{8(PS)}\alpha_{V(PS)},  
\end{equation}

\begin{equation}
  g_{\eta\eta\rho} = 0,  
\end{equation}

\begin{eqnarray}
  g_{KK\omega} = g_{1(PS)}\cos\theta_{V} \nonumber \\ + g_{8(PS)}\sin\theta_{V} \frac{1}{3}\sqrt{3}(4\alpha_{V(PS)} -1),   \label{komega}
\end{eqnarray}

\begin{eqnarray}
  g_{\bar{K}\bar{K}\omega} = g_{1(PS)}\cos\theta_V \nonumber \\   - g_{8(PS)}\sin\theta_V \frac{1}{3}\sqrt{3}(1 + 2\alpha_{V(PS)}).  \label{kbaromega}
\end{eqnarray}

\begin{eqnarray}
  g_{\eta\eta\omega} = g_{1(PS)}\cos\theta_{V} \nonumber \\ - g_{8(PS)}\sin\theta_{V} \frac{2}{3}\sqrt{3}(1 - \alpha_{V(PS)}),  
\end{eqnarray}

\begin{eqnarray}
  g_{\pi\pi\omega} =  g_{1(PS)}\cos\theta_{V} \nonumber \\ + g_{8(PS)}\sin\theta_{V} \frac{2}{3}\sqrt{3}(1 - \alpha_{V(PS)}),  
\end{eqnarray}
and again, the results for the $\phi$ meson is obtained by replacing $\sin\theta_V \to \cos\theta_V$ and $\cos\theta_V \to  -\sin\theta_V$. The subscript $(PS)$ is used to indicate that these results are related to the pseudoscalar octet instead of the baryonic one. Moreover, as the mixing angle $\theta_V$ is related solely to the nature of the $\phi$ and $\omega$ meson, once more we have $\theta_V = 35.264º$.

Now, the main difference between the two octets is that while in the baryonic octet each member is a different baryon, in the pseudoscalar mesonic octet, we have particles and antiparticles. Therefore, we can use the G-Parity~\cite{GreinerSymmetries} to help us fix the three free parameters. 

The G-Parity gives the interaction of an anti-particle with a determined meson as a function of the interaction of the particle with the same meson. For a mesonic field $M$, we have:

\begin{equation}
  G|M\rangle = \pm |M\rangle  
\end{equation}

First, the $\rho$ meson has a positive G-Parity~\cite{GreinerSymmetries}, which implies:

\begin{equation}
g_{\bar{K}\bar{K}\rho} =  g_{KK\rho}. 
\end{equation}

Comparing Eqs.~\ref{krho} and~~\ref{kbarrho}, we see that it can only be solved by imposing $\alpha_{V(PS)} = 1.00$.

On the other hand, the $\omega$ and $\phi$  mesons have a negative G-Parity, impling:

\begin{equation}
 g_{\bar{K}\bar{K}\omega} =  - g_{KK\omega}; \quad    g_{\bar{K}\bar{K}\phi} = - g_{KK\phi}. \label{Gomega}
\end{equation}

With $\alpha_{V(PS)}$ already set to one, by comparing Eqs.~\ref{komega} and~\ref{kbaromega}, it can be seen that Eq.~\ref{Gomega} results in $g_{1(PS)} = 0$, or $z_{(PS)}^{-1} = 0$;
which also implies $g{\pi\pi\omega} =g_{\eta\eta\omega} = 0$.
Another desirable result obtained with this approach is $g_{\pi\pi\phi} = 0$.

With those definitions, the strength of the kaon interaction concerning the nucleonic one is:

\begin{equation}
    \frac{g_{KK\omega}}{g_{NN\omega}} =  \frac{g_{8(PS)}}{g_8} \bigg (\frac{3}{5 +4\alpha_V} \bigg ); \quad \frac{g_{KK\rho}}{g_{NN\rho}} =  \frac{g_{8(PS)}}{g_8}.
\end{equation}

Now, in the SU(6) limit ($\alpha_V \to 1.00)$, the naive quark-isospin counting rule is restored for the members of the baryon octet. Assuming this is also true for the members of the pseudoscalar meson octet implies:

\begin{equation}
 \frac{g_{8(PS)}}{g_8} = 1.00.
 \end{equation}

Consequently, the coupling of (anti)kaons, as well as the coupling of the hyperons with the vector meson octet, depends on a single parameter, $\alpha_V$:

\begin{eqnarray}
     \frac{g_{KK\omega}}{g_{NN\omega}} =  \frac{3}{5 +4\alpha_V}; \quad \frac{g_{KK\rho}}{g_{NN\rho}} = 1; \quad \frac{g_{KK\phi}}{g_{KK\omega}} = \sqrt{2}.\nonumber \\
\end{eqnarray}

As anti-kaons can potentially condense in neutron stars' interiors, it is worth discussing the role of each meson in both anti-kaon-anti-kaon interaction and anti-kaon-nucleon interaction. 

\begin{itemize}
    \item $\omega$ meson: $\omega$ meson is a repulsive channel in anti-kaon-anti-kaon interaction. However, as $g_{\bar{K}\bar{K}\omega}$ is negative, while $g_{NN\omega}$ is positive, this means that the anti-kaon-nucleon interaction is attractive (analogous to the role played by the $\rho$ meson in $p$-$p$,$n$-$n$ and $p$-$n$ interactions). In the same sense, as $g_{KK\omega}$ is also positive, the kaon-nucleon interaction is repulsive, a result experimentally confirmed~\cite{LI1997}.

    \item $\rho$ meson: $\rho$ meson is a repulsive channel. The coupling constant is positive for kaons, anti-kaons, and nucleons. However, $\tau_3$ reads +1 to protons, $K^+$ and $\bar{K}^0$; and reads  -1 to neutrons, $K^0$ and $K^-$.

    \item $\phi$ meson: $\phi$ meson is a repulsive channel between anti-kaons. It does not couple to the nucleons and has a negative coupling to the hyperons. This means that the $\phi$ meson is also repulsive in anti-kaon-hyperon interaction, while kaon-hyperon interaction is attractive in this channel. A desirable result, as both hyperons and anti-kaons have an $s$ quark, while kaons have an $\bar{s}$ one.

    \item $\sigma$ meson: with a positive G-Parity,  the scalar $\sigma$ meson is always an attractive channel.
\end{itemize}

The use of G-Parity gives us a clear picture of the role of each meson, reproducing the negative sign usually found in the anti-kaon potential depth (Eq.~\ref{mesonpd}) and dispersion relation (Eq.~\ref{disprel})~\cite{kaonp,Banik2001}, but in a more organic way.

{ It is worth to pointing out that the SU(3) and SU(6) symmetry are ultimattely broken due to the large $s$-quark mass. However, the effects are very small due to the Ademollo-Gatto theorem~\cite{Ademollo1964} that prevents first-order symmetry-breaking effects for the vector coupling constants $g's$. Indeed, an explicit calculation about the symmetry breaking effects was done in ref.~\cite{RATCLIFFE1996}, where the author shows that these effects are lower than 0.2$\%$. }

Now, to describe the nucleon-nucleon interaction, I use the eL3$\omega\rho$ parametrization of the QHD~\cite{lopesPRD}. This parametrization virtually satisfies all constraints at the saturation point for the symmetric nuclear matter~\cite{Dutra2014,Micaela2017,Essick2021}, as well as constraints related to neutron stars' observations~\cite{Riley2021,Miller2021,AbbottPRL}. The parameters of the model, the predicted values of six physical quantities at the saturation density, and the phenomenological constraints are presented in Tab.~\ref{TL1}.

 \begin{table}[t]
     \centering
     \begin{tabular}{c|c}
     \hline
     \multicolumn{2}{c}{eL3$\omega\rho$}
     \\ \hline
    $(g_{NN\sigma}/m_s)^2$ & 12.108 fm$^2$ \\
    $(g_{NN\omega}/m_v)^2$ & 7.132  fm$^2$ \\
    $(g_{NN\rho}/m_\rho)^2$ & 5.85  fm$^2$ \\
    $\kappa$ & 0.004138 \\
    $\lambda$ &  $-0.00390$ \\
    $\Lambda_{\omega\rho}$ &  0.0283 \\ 
   \hline
     \end{tabular}
     
\begin{tabular}{cc}
    { }&{ } \\
\end{tabular}

\begin{tabular}{c|cc}
\hline 
Quantity & Constraint & This model\\\hline
$n_0$ ($fm^{-3}$) & 0.148--0.170 & 0.156 \\
   $M^{*}/M$ & 0.60--0.80 & 0.69  \\

  $K$ (MeV)& 220--260  &  256  \\

 $S_0$ (MeV) & 31.2--35.0 &  32.1  \\
$L$ (MeV) & 38--67 & 66\\

 $B/A$ (MeV) & 15.8--16.5  & 16.2  \\ 
 \hline
\end{tabular}
 
\caption{ The eL3$\omega\rho$~\cite{lopesPRD} parameterization (top) and its predictions for nuclear matter (bottom). The phenomenological constraints are taken from Ref.~\cite{Dutra2014, Micaela2017, Essick2021}. } 
\label{TL1}
\end{table}

\begin{table}[t!]
\begin{center}
\begin{tabular}{ccccc}
\hline 
  $\alpha_V = $ & SU(6) & $0.75$ & 0.50 & 0.25  \\
 \hline
 $g_{\Lambda\Lambda\omega}/g_{NN\omega}$ & 0.667 & 0.687 & 0.714 & 0.750 \\
 $g_{\Sigma\Sigma\omega}/g_{NN\omega}$ & 0.667  & 0.812 & 1.00 & 1.25   \\
 $g_{\Xi\Xi\omega}/g_{NN\omega}$ & 0.333 & 0.437 & 0.571 & 0.750  \\
 $g_{\Lambda\Lambda\phi}/g_{NN\omega}$ &  -0.471  & -0.619 & -0.808 & -1.06  \\
 $g_{\Sigma\Sigma\phi}/g_{NN\omega}$ & -0.471 & -0.441 & -0.404 & -0.354 \\
 $g_{\Xi\Xi\phi}/g_{NN\omega}$ & -0.943 & -0.972 & -1.01 & -1.06 \\
 $g_{\Lambda\Lambda\rho}/g_{NN\rho}$ & 0.0 &  0.0 & 0.0 & 0.0 \\
 $g_{\Sigma\Sigma\rho}/g_{NN\rho}$ & 2.0 & 1.5 & 1.0 & 0.5 \\
 $g_{\Xi\Xi\rho}/g_{NN\rho}$ & 1.0 & 0.5 & 0.0 & -0.5 \\
 $g_{\Sigma\Lambda\rho}/g_{NN\rho}$ & 0.00 & 0.288 & 0.577 & 0.866 \\
 $g_{\bar{K}\bar{K}\omega}/g_{NN\omega}$ & -0.333 & -0.375 & -0.429 & -0.50\\ 
 $g_{\bar{K}\bar{K}\phi}/g_{NN\omega}$  & -0.471 & -0.530 & -0.607 & -0.707 \\
 $g_{\bar{K}\bar{K}\rho}/g_{NN\rho}$  & 1.0 & 1.0 & 1.0 & 1.0 \\
 $g_{\Lambda\Lambda\sigma}/g_{NN\sigma}$ & 0.610 & 0.625 & 0.646 & 0.674 \\
 $g_{\Sigma\Sigma\sigma}/g_{NN\sigma}$ & 0.406  & 0.518 & 0.663 & 0.855   \\
 $g_{\Xi\Xi\sigma}/g_{NN\sigma}$ & 0.269 & 0.350 & 0.453 & 0.590   \\
 $g_{\bar{K}\bar{K}\sigma}/g_{NN\sigma}$ & 0.234 & 0.201 & 0.160  & 0.105\\  
\hline
 $U_K(n_0)$ (MeV) & +6  & +25 & + 48 & +80  \\
\hline 
\end{tabular}
 
\caption{Coupling constants of hyperons and anti-kaons for different values of $\alpha_V$. The kaon potential depth is also presented.} 
\label{T2}
\end{center}
\end{table}

The interaction of the hyperons and anti-kaons with the scalar meson is fixed in such a way that it reproduces their potential depths. For the hyperons~\cite{Potentials2000,LQCD}: $U_{\Lambda} = -28$ MeV, $U_{\Sigma}$ = +30 MeV, and $U_\Xi = -4$ MeV. The anti-kaons do not have such a degree of precision. As pointed out in Ref.~\cite {Thapa2021}, different models and calculations yield different values for the potential depth. Acceptable values lie around $- 200$ MeV$~<U_{\bar{K}}~<$ $- 40$ MeV.  Here, I use $U_{\bar{K}} = -140$ MeV, a value usually found in the literature.
Thanks to G-Parity, I can theoretically predict the kaon potential depth. The coupling constants and the kaon potential depth as a function of $\alpha_V$ are presented in Tab.~\ref{T2}. It is also worth pointing out that the couplings for both kaons and hyperons with the vector mesons are fully model-independent.

\section{Numerical Results}\label{mr}

%%%%%%%%%%%%%%%%%
\begin{figure*}[h!]
\begin{tabular}{ccc}
\centering % \begin{center}/\end{center} takes some additional vertical space
\includegraphics[scale=.52, angle=270]{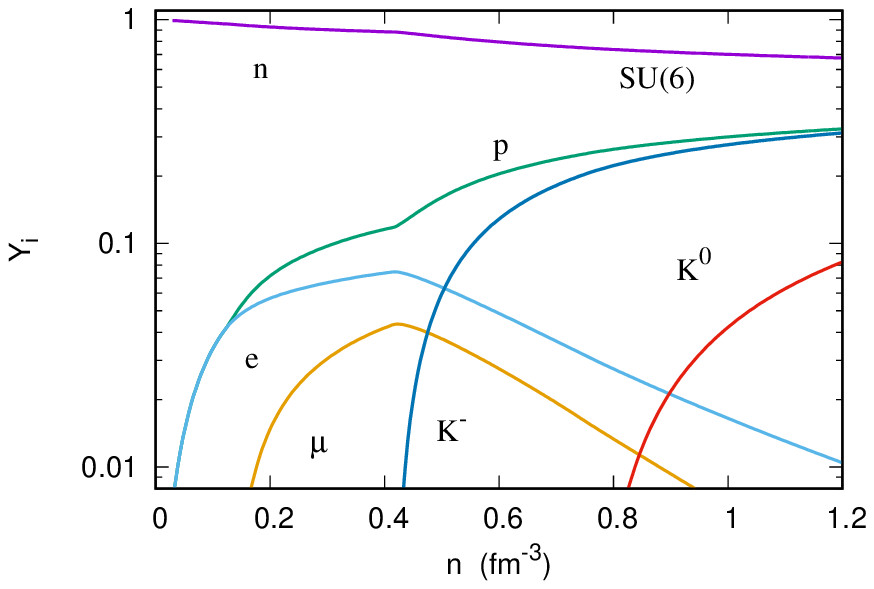} &
\includegraphics[scale=.52, angle=270]{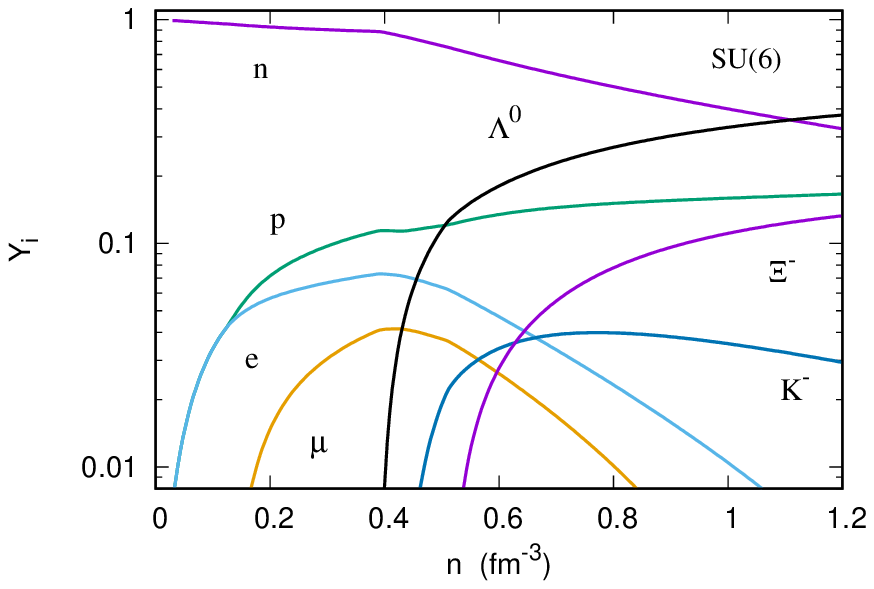} \\
\includegraphics[scale=.52, angle=270]{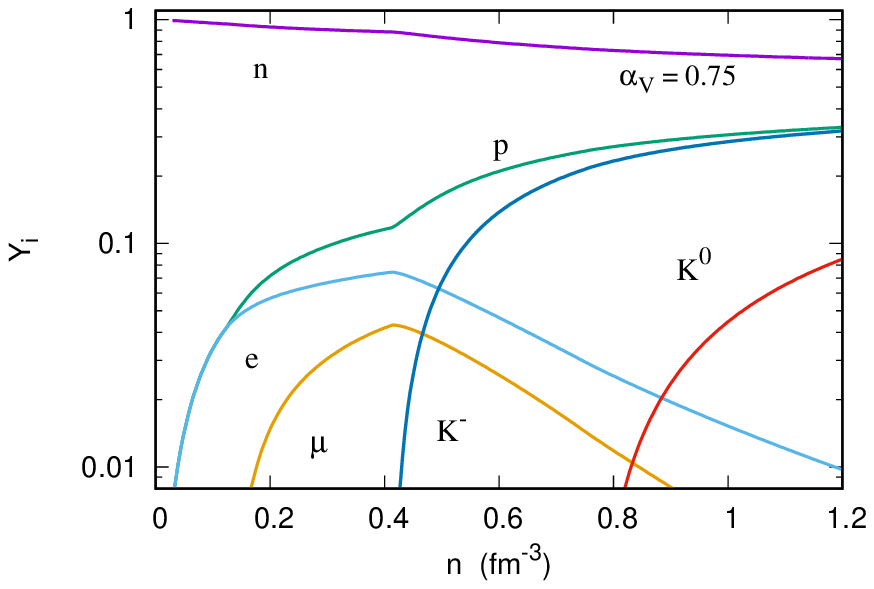} &
\includegraphics[scale=.52, angle=270]{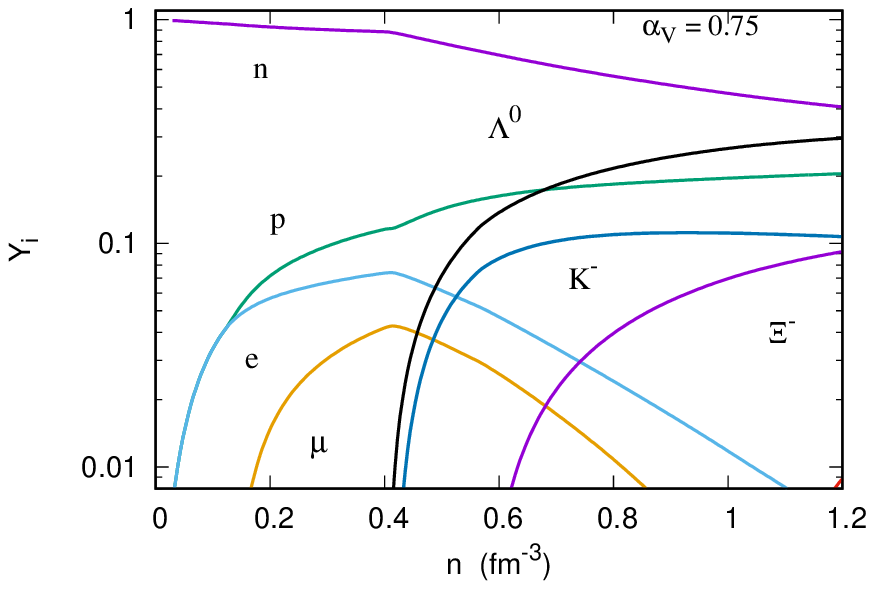}\\
\includegraphics[scale=.52, angle=270]{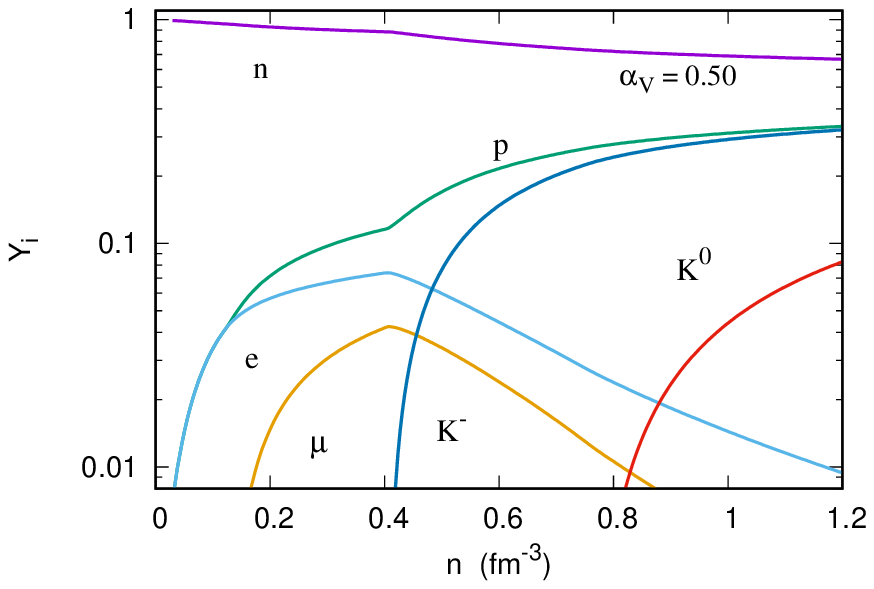} &
\includegraphics[scale=.52, angle=270]{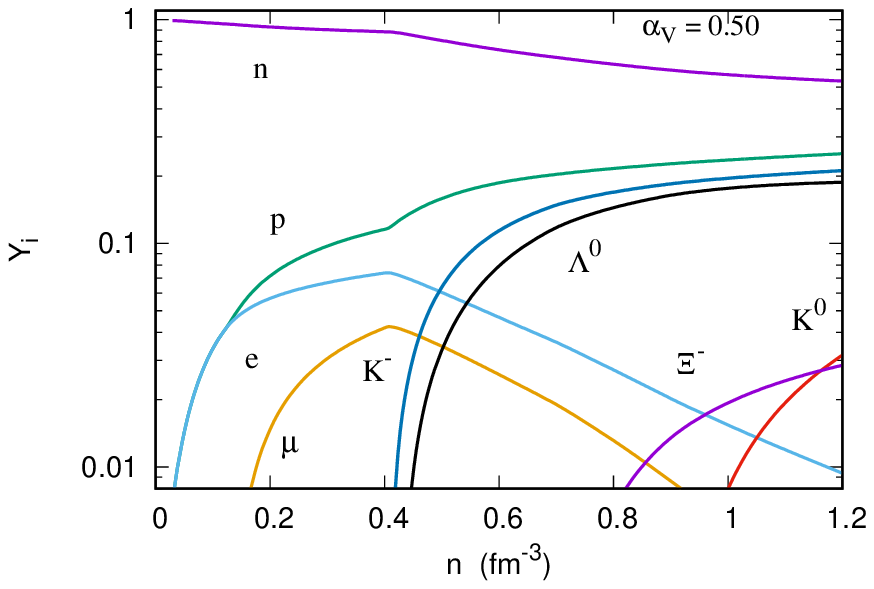}\\
\includegraphics[scale=.52, angle=270]{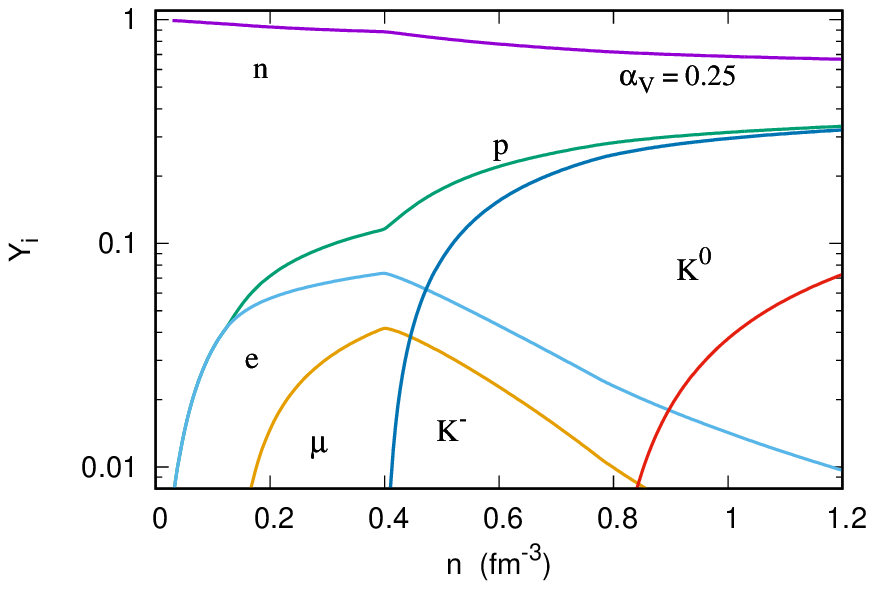} &
\includegraphics[scale=.52, angle=270]{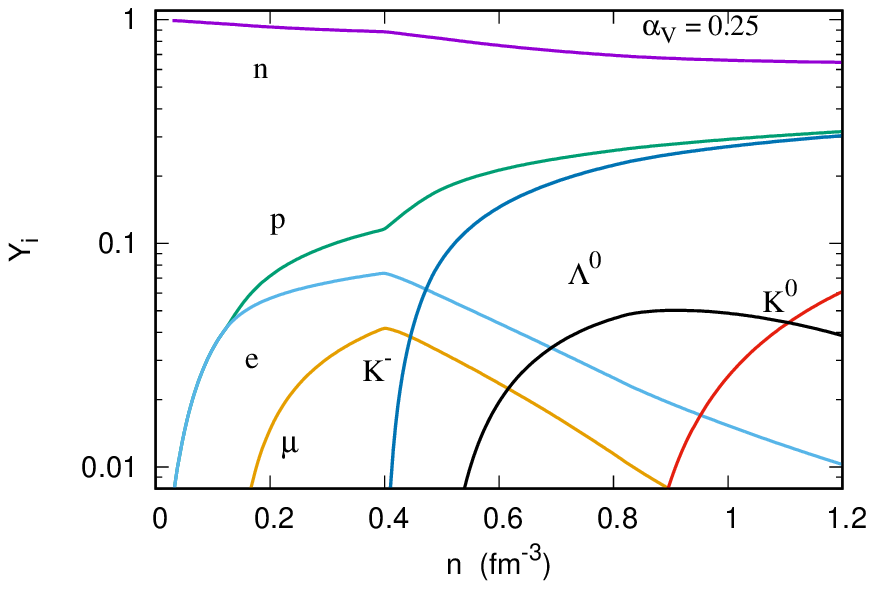}\\
\end{tabular}
\caption{Particle population for different values of $\alpha_V$ considering nucleons and anti-kaons (left) and nucleons, anti-kaons, and hyperons (right).} \label{F1}
\end{figure*}

I start the analysis by pointing out that as we move away from SU(6) by reducing the value of $\alpha_V$, the absolute value of $g_{\bar{K}\bar{K}\omega}$ increases, indicating a strong nucleon-$\bar{K}$ attraction at high densities. However, this also causes an increase in the repulsive $\phi$ channel. There is a competition between these two channels, but as the $\phi$ meson does not couple to the nucleons, the EOS will become softer as we reduce $\alpha_V$.

A second point is the potential depth for the kaons at the saturation density, $U_K (n_0)$. In the SU(6) limit, this potential is only weakly repulsive, $U_K = +6$ MeV, but it becomes strongly repulsive for $\alpha_V = 0.25$. $U_K = +80$ MeV. In the future, measurements of the kaon potential depth can help us to constrain the values of $\alpha_V$.

In Fig.~\ref{F1}, I present the particle population as a function of $\alpha_V$ for two different approaches: on the left, the matter is composed of nucleons+anti-kaons, while on the right, there are nucleons+hyperons+anti-kaons.  In the first case, reducing the value of $\alpha_V$ causes only a small increase in the anti-kaons on the nuclear matter. This can be understood due to two other mesons, the $\sigma$ and the $\phi$. The reduction of $\alpha_V$ leads to a increase on the absolute value of attractive coupling  $g_{\bar{K}\bar{K}\omega}$, which reduces the other attractive channel, $g_{\bar{K}\bar{K}\sigma}$ in order to keep the potential depth $U_{\bar{K}} = -140$ MeV constant. Moreover, the reduction of $\alpha_V$ also increases the absolute value of $g_{\bar{K}\bar{K}\phi}$, which is a repulsive channel between the anti-kaons.

The situation is much more interesting in the case of nucleons+hyperons+anti-kaons.
The reduction of $\alpha_V$ leads to a simultaneous increase in the absolute value of  $g_{\bar{K}\bar{K}\omega}$ and  $g_{YY\omega}$. However, the negative sign of $g_{\bar{K}\bar{K}\omega}$  makes the anti-kaons become more and more energetically favorable, while the positive sign in  $g_{YY\omega}$ causes hyperons to become increasingly energetically unfavorable. For instance, in the SU(6) parametrization, the relatively small values of the coupling constants lead to a significant reduction of the $K^-$ while the $\bar{K}^0$ was totally suppressed. As we move apart from the SU(6), the absolute values of the coupling constants increase, and the fraction of the hyperons is reduced at the same time that the fraction of $K^-$ increases. For $\alpha_V = 0.50$, the $K^-$ becomes the first non-nucleonic degree of freedom to appear in dense matter, while the $\bar{K}^0$ appears at high densities. 
For $\alpha_V = 0.25$, there is a total suppression of the $\Xi^-$, and a strong reduction of $\Lambda^0$ hyperons. This reduction is also strongly influenced by the mixed term $g_{\Sigma\Lambda\rho}$~\cite{lopes2023ptep}. The reader interested in the effects of $\alpha_V$ matter without kaons is referred to Fig. 1 of ref.~\cite{lopesPRD}.

%%%%%%%%%%%%%%%%%
\begin{figure*}[ht]
\begin{tabular}{ccc}
\centering % \begin{center}/\end{center} takes some additional vertical space
\includegraphics[scale=.52, angle=270]{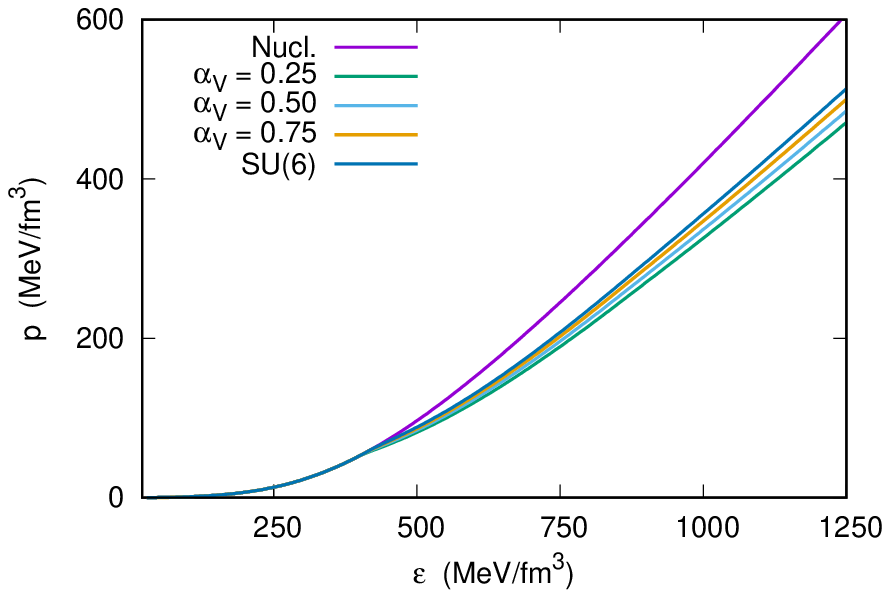} &
\includegraphics[scale=.52, angle=270]{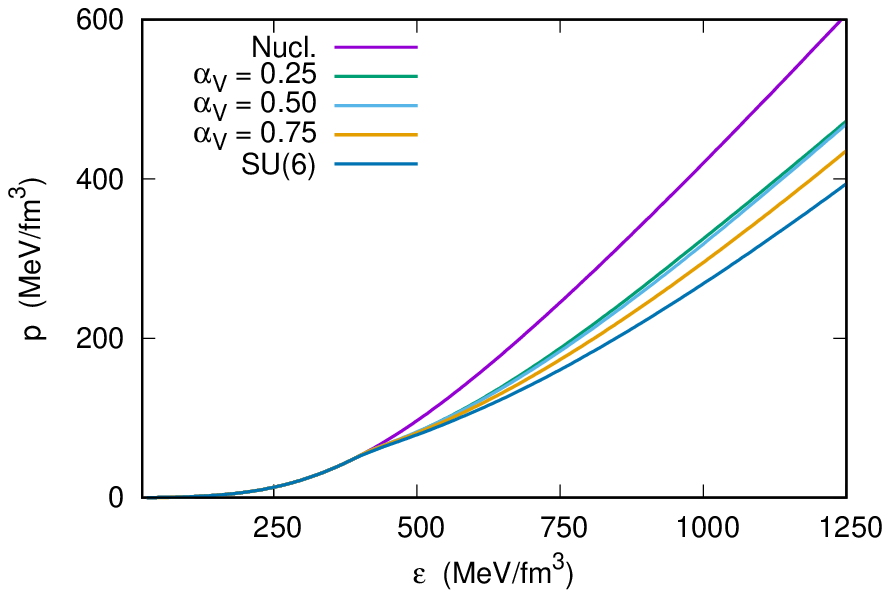} \\
\includegraphics[scale=.52, angle=270]{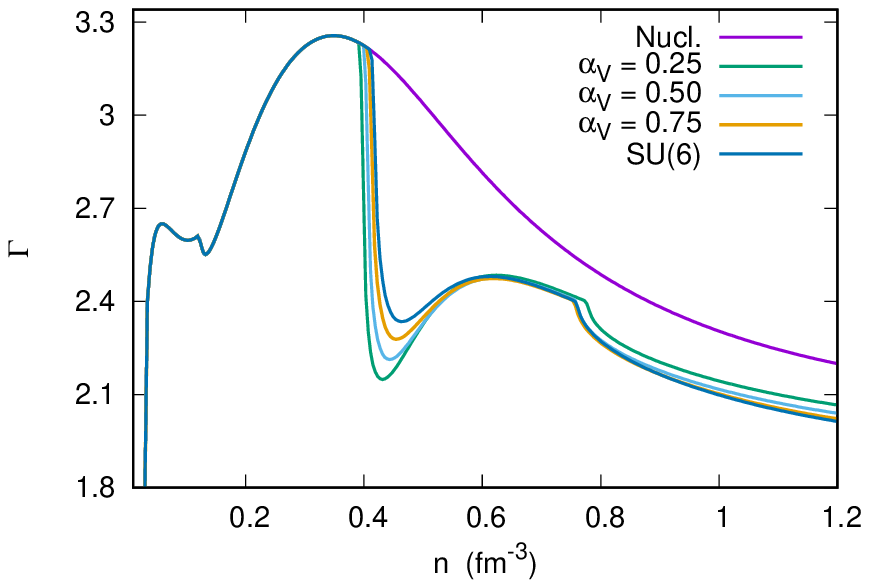} &
\includegraphics[scale=.52, angle=270]{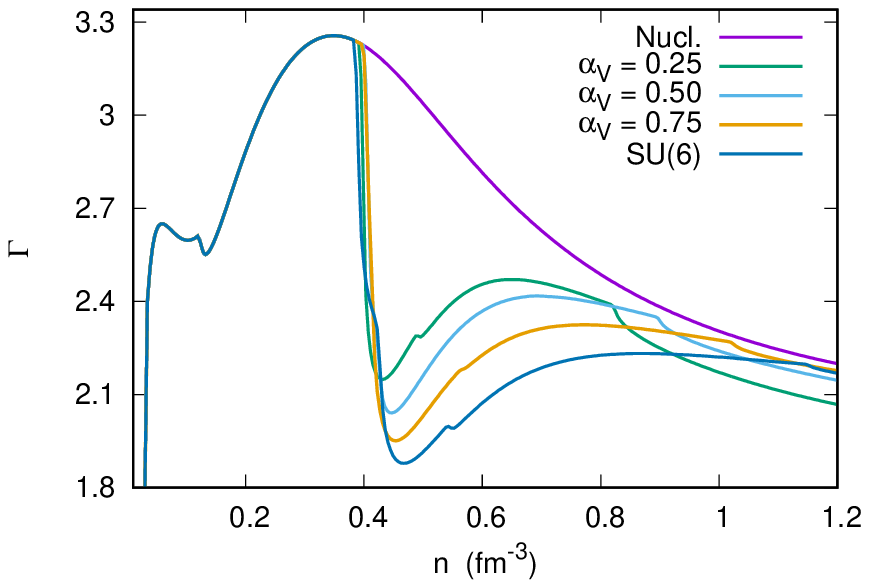} \\
\includegraphics[scale=.52, angle=270]{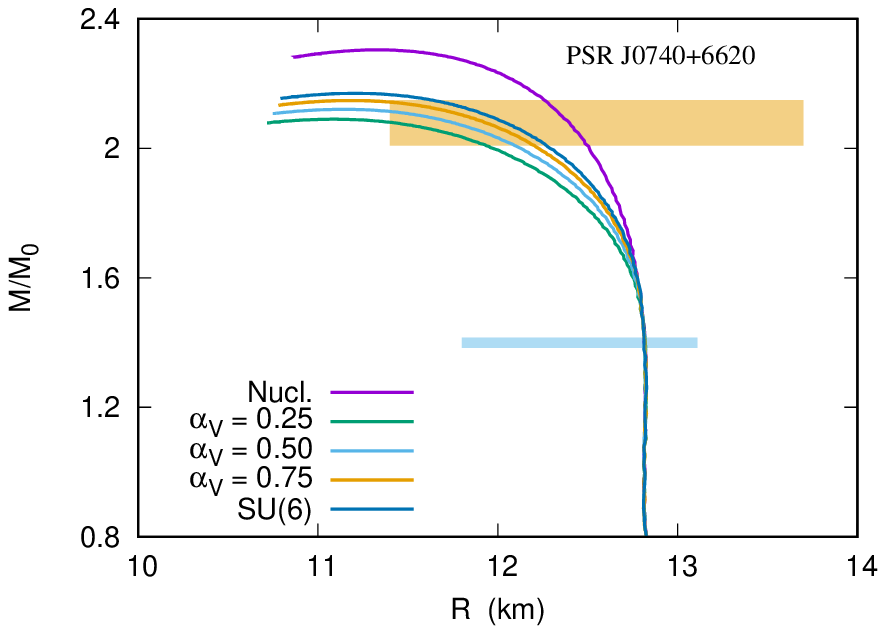} &
\includegraphics[scale=.52, angle=270]{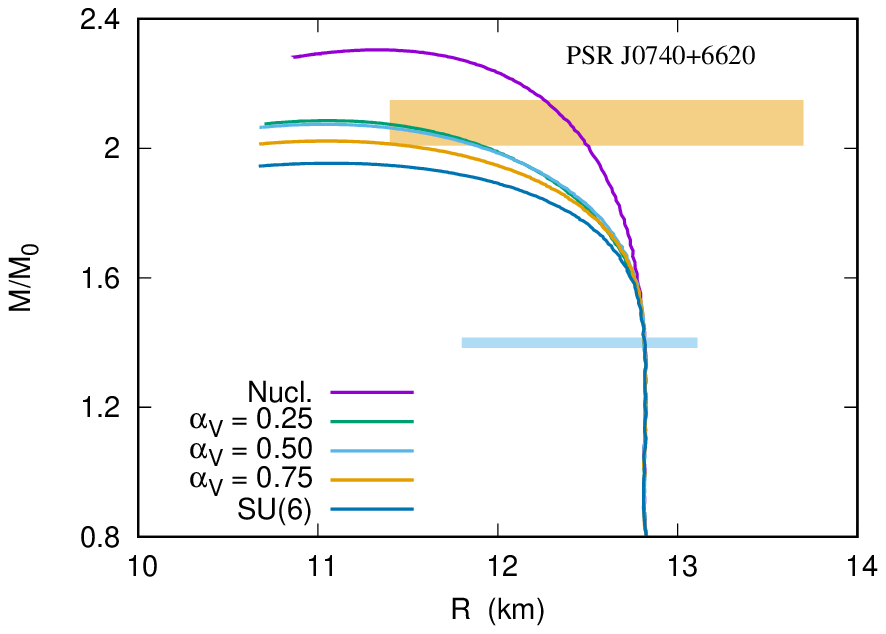} \\
\end{tabular}
\caption{EOSs (top), adiabatic index (middle), and TOV solutions (bottom) within different values of $\alpha_V$. Figures on the left represent nucleons+anti-kaons matter, while figures on the right means nucleons+hyperons+anti-kaons.} \label{F2}
\end{figure*}

Now we turn our attention to the equation of state (EOS). In the top of Fig.~\ref{F2} I plot the EOSs for nucleons+anti-kaons on the left and nucleons+hyperons+anti-kaons on the right for different values of $\alpha_V$. For nucleons+anti-kaons, it was shown that as we move away from SU(6), we only obtain a small increase in anti-kaons. As anti-kaons do not produce pressure in S-wave condensate, a larger amount of anti-kaons produces softer EOSs. The softening of the EOS due to the presence of anti-kaons is easily recognized, but the differences due to different values of $\alpha_V$ are small.

Here again, the situation is more interesting when we are dealing with nucleons+hyperons+anti-kaons. It is a well-known effect that reducing the value of $\alpha_V$ stiffens the EOS for nucleons+hyperon matter~\cite{WeissPRC2012,lopesPRD}. However, as lowering the value of $\alpha_V$ favors the onset of anti-kaons, the stiffening of the EOS is severely compromised.

A microscopic quantity that is very sensitive to the onset of new degrees of freedom is the adiabatic index, $\Gamma$:

\begin{equation}
 \Gamma = \frac{(p+\epsilon)}{p} \bigg (\frac{\partial p}{\partial \epsilon} \bigg ). \label{add}
\end{equation}
For multicomponent matter, the adiabatic index exhibits jumps at densities coincident with density thresholds of individual components, signaling phase transitions, and/or changes in the matter constitution~\cite{lopesPRD,Haensel2002}. The adiabatic index, $\Gamma$, presents information not only on the EOS (p and $\epsilon$) but also on the speed of sound. The behavior of $\Gamma$ for nucleons+anti-kaon (left) and nucleons+hyperons+anti-kaons (right) matter are displayed at the middle of Fig.~\ref{F2} for different values of $\alpha_V$.

On the left, we see that the onset of the $K^-$ causes a drop in the adiabatic index around 0.41 fm$^{-3}$. There is a subsequent increase followed by a new drop around 0.75 fm$^{-3}$ due to the onset of the $\bar{K}^0$. Qualitatively, all the curves present the same behavior. Quantitatively, as lower values of $\alpha_V$ produce a larger amount of anti-kaon, the height of the drop due to the onset of $K^-$ is larger for lower values of $\alpha_V$.

On the right side, besides the quantitative, there are also qualitative differences due to the competition of anti-kaons and hyperons. In the SU(6) limit, the anti-kaons are strongly suppressed. and the adiabatic index is very close to the one for nucleons+hyperons matter (see Fig. 2 of ref.~\cite{lopesPRD}).  But, as we reduce the value of $\alpha_V$, the anti-kaons become more and more populated while at the same time, the fraction of hyperons is reduced. Therefore, each curve for $\Gamma$ is unique even in shape. For $\alpha_V = 0.25$, the hyperon suppression is strong enough that the $\Gamma$ of nucleons+hyperons+anti-kaons is very similar to the $\Gamma$ of nucleons+anti-kaons.

Now, I discuss the macroscopic features of neutron stars by using the EOSs as an input to solve the TOV equations~\cite{TOV}:

\begin{eqnarray}
 \frac{dp}{dr} = \frac{-GM(r)\epsilon (r)}{r^{2}} \bigg [ 1 + \frac{p(r)}{\epsilon(r)} \bigg ]   \bigg  [ 1 + \frac{4\pi p(r)r^3}{M(r)} \bigg ] \nonumber \\ \times \bigg [ 1 - \frac{2GM(r)}{r} \bigg ]^{-1} , \nonumber \\
 \frac{dM}{dr} =  4\pi r^2 \epsilon(r).  \nonumber \\  \label{Etov}
\end{eqnarray}

The mass-radius relations are presented at the bottom of Fig.~\ref{F2} for nucleons+anti-kaons (left) and nucleons+hyperons+anti-kaons(right). Altogether, I also display two observational constraints obtained by the NICER X-ray telescope~\cite{Miller2021,Riley2021}. Nowadays, maybe the more important astrophysical constraint is the PSR J0740+6620 pulsar, whose values of mass and radius are $M = 2.08  \pm 0.07 M_\odot$ and $R= 12.39^{+1.30}_{-0.98}$ km, respectively. Any realistic EOS must be able to fulfill this constraint.
Another important constraint is the radius of the canonical $M  = 1.4 M_\odot$ star. Ref.~\cite{Miller2021},  suggests $R_{1.4} = 12.45 \pm  0.65$ km. This limits the radius of the canonical star within an uncertainty of only 5\%. Furthermore, to describe the outer and inner crusts of the neutron stars, we utilized the Baym-Pethick-Sutherland (BPS) EOS \cite{BPS} and the Baym-Bethe-Pethick (BBP) EOS \cite{BBP}, respectively.   %We use the BPS+BBP EoS up to the density of 0.0089 fm$^{-3}$ for all values of $k$, and from this point on, we use the QHD EOS, as suggested in ref.~\cite{Glenbook}. This procedure is the same as the one done in ref.~\cite{lopesPRD,lopes2024PRCb,lopesnpa,lopes2023ptep,Lopes2022ApJ}, .

As can be seen, for the nucleons+anti-kaons, the presence of anti-kaons reduces the maximum mass, although the results are only weakly dependent on the value of $\alpha_V$. While for a pure nucleonic star, the maximum mass is  2.30 $M_\odot$, for the SU(6), we have $M = 2.17M_\odot$ and for $\alpha_V = 0.25$, $M  =2.09M_\odot$. As expected, moving away from the SU(6) reduces the maximum mass. Furthermore, all values of $\alpha_V$ satisfy the constraint related to the PSR J0740+6620 pulsar.

For nucleons+hyperon+anti-kaons, the situation is reversed: move away from SU(6) stiffen the EOSs and produce more massive neutron stars. However, the stiffening of the EOS is strongly compromised. Without the (anti)kaons, it was shown in ref.~\cite{lopesPRD} (see Fig. 4), that moving from the SU(6) to $\alpha_V = 0.25$ increases the maximum mass from 1.97 $M_\odot$ to 2.20 $M_\odot$. However, the maximum mass here reaches only 2.08 $M_\odot$ for $\alpha_V = 0.25$, very close to the case of nucleons+anti-kaons matter. The constraint related to the PSR J0740+6620 pulsar is only obtained for $\alpha_V = 0.75$, $\alpha_V = 0.50$ and $\alpha_V$ = 0.25.

The lowest neutron star mass that presents a non-nucleonic degree of freedom lies between 1.43 - 1.48$M_\odot$, depending on the model. This implies that all 1.4$M_\odot$ stars present the same radius of 12.82 km and satisfy the constraint presented in ref.~\cite{Miller2021}, $R_{1.4} = 12.45 \pm  0.65$ km.

Another constraint satisfied by all models is related to the dimensionless tidal parameter, $\Lambda$~\cite{Hinderer_2008,Chat2020}. It was pointed out in ref.~\cite{AbbottPRL} that for the canonical star, $70~<\Lambda_{1.4}~<580$. As can be seen in Fig. 5 of ref.~\cite{lopesPRD}, the eL3$\omega\rho$ model predicts $\Lambda_{1.4}$ = 516, in agreement with this constraint. The main results for nucleonic+anti-kaons $(N\bar{K})$ and nucleons+hyperons+anti-kaons $(NY\bar{K})$ are summarized in Tab.~\ref{T3}.

\begin{table}[t!]
\begin{center}
\begin{tabular}{ccccc}
\hline
   & $\alpha_V$ &$M_{\mathrm{max}} (M_\odot)$ &  $n_c$ (fm$^{-3}$) & $R_{max}$ (km)  \\
\toprule
 $N$ & - & 2.30 &  1.04 &11.34  \\
 $N\bar{K}$ & SU(6) & 2.17 & 0.94 &11.21  \\
 $NY\bar{K}$& SU(6) & 1.95 & 1.06 & 11.08  \\ 
$N\bar{K}$& 0.75 & 2.15 & 0.97 &11.19  \\
$NY\bar{K}$& 0.75 &2.02& 1.05 & 11.06  \\
  $N\bar{K}$& 0.50 &2.12 &  0.99 & 11.14  \\
 $NY\bar{K}$&0.50 &2.07 & 1.03 & 11.04 \\
 $N\bar{K}$&0.25 &2.09 & 1.02 & 11.07   \\
 $NY\bar{K}$&0.25 &2.08 & 1.03 & 11.06  \\

\hline
\end{tabular}
\caption{Neutron stars' main properties for different values of $\alpha_V$. All models predict $R_{1.4} = 12.82$ km and $\Lambda_{1.4}$ = 516.} \label{T3}
\end{center}
\end{table}

\subsection{Additional Constraints}

{ As pointed out earlier, the eL3$\omega\rho$ satisfies all nuclear constraints coming from ref.~\cite{Dutra2014,Micaela2017,Essick2021} presented in Tab.~\ref{TL1}, as well as the constraints related to  PSR J0740+6620~\cite{Miller2021,Riley2021} and the canonical 1.4$M_\odot$ coming from both, the radius inferred by the NICER X-ray telescope~\cite{Miller2021}, and the dimensionless tidal parameter from the LIGO/VIRGO gravitational wave observatories~\cite{AbbottPRL}. Nevertheless, it is worth discussing more recent additional constraints related to both, nuclear and astrophysical sources. First, in ref.~\cite{HuthNature2022}, combining chiral effective field theory (CEFT), heavy-ion collision (HIC) and astrophysical results coming from the NICER X-ray telescope, the authors constrain the pressure of beta-stable matter at 1.5$n_0$ and the radius of the canonical star. The eL3$\omega\rho$ predicts $p(1.5n_0) = 8.72$ MeV/fm$^{3}$ and $R_{1.4} = 12.82$ km. Using CEFT up to $1n_0$, the authors constrain $p(1.5n_0) = 9.12^{+6.66}_{-4.36}$ MeV/fm$^{3}$ and $R_{1.4} = 12.56^{+1.07}_{-1.01}$ km. In this case, the eL3$\omega\rho$ satisfies both constraints. However, by using CEFT up to $1.5n_0$ the authors find  $p(1.5n_0) = 6.25^{+1.90}_{-2.26}$ MeV/fm$^{3}$ and $R_{1.4} = 12.01^{+0.78}_{-0.77}$ km. This implies that the values of eL3$\omega\rho$  are above the upper limit by $\approx~7\%$ in the case $p(1.5n_0)$,  and $\approx~0.3\%$ for the radius of the canonical star. Therefore, the CEFT predicts EOSs softer than the presented model of the QHD.

Astronomical observation also provides us with additional constraints from other pulsars alongside the already discussed PSR J0740+620. An example is the PSR J0030+0451, which inferred mass and radius of $1.34^{+0.15}_{-0.16}M_\odot$ and $12.71^{+1.14}_{-1.19}$ km respectivally~\cite{Riley:2019yda}. Such a constraint is satisfied by the eL3$\omega\rho$ model. On the other hand, the PSR J0437–4715 has an inferred mass of 1.418 $\pm$ 0.037 $M_\odot$ and a radius of only $11.36^{+0.95}_{-0.63}$ km~\cite{Choudhury_2024}. In this case, the presented model is outside the upper limit by $\approx$ 4$\%$. A much more extreme case is the HESS J1731-347. With a mass of only 0.77$M_\odot$ and a radius lying in the range 10.4$^{+0.86}_{-0.78}$ km~\cite{Doroshenko_2022}, not even its true nature is well understood, but some studies point to it being a strange star~\cite{lopesEPJC2025}.}

{ Before the conclusions, the attentive reader may wonder if other hadrons can also play a role in neutron stars' interior. Pions seem to be the more logical choice, as they are the lightest of the pseudoscalar mesons. However, since the early 1980s, we have known that the pion condensate is strongly suppressed. In ref.~\cite{DICKHOFF1983}, the authors show that pion self-energy yields a considerable repulsion in the pionic sector. More modern calculations show that the repulsive pion potential depth can be as large as +100 MeV~\cite{OhnishiPRC2009}. 

A more interesting case is the possible presence of $\Delta$ resonances. Indeed, some studies indicate that a large amount of $\Delta's$ can be present, causing a strong influence on neutron stars' macroscopic properties~\cite{Kauan2022PRC,lopesPRD}. Notwithstanding, the possible presence of $\Delta's$ in nuclear matter has also faced some criticism. For instance, in ref.~\cite{BETHE1974}, the authors discuss whether the $\Delta's$ can be treated as point-like particles like the nucleons and hyperons, or whether they are intrinsically large and complex structures. Using QMC models, ref.~\cite{MOTTA2020} has shown that $\Delta$ particles are never present. As can be seen, the presence of $\Delta's$ is a rich but complex subject. Such a study is left for future work.}

\section{Final Remarks}\label{fr}

In this work, I calculated the coupling constants for both the baryon octet and the pseudoscalar meson octet with the members of the vector meson octet in a unified broken SU(6) symmetry. Within this approach, all coupling constants related to the vector mesons are left as a function of only one free parameter, $\alpha_V$.  The main conclusions are summarized below.

\begin{itemize}

      \item The members of the baryon octet and the members of the pseudoscalar meson octet present the same quantum numbers and therefore have the same CG coefficients. However, while each member of the baryon octet is an independent baryon, the pseudoscalar baryon octet contains particles and antiparticles. This allows us to use the G-Parity to eliminate the free parameters of the SU(3) group.

    \item With the combined use of the symmetry group arguments and G-parity, several expected results are naturally obtained, for instance $g_{\bar{K}\bar{K}\omega} = -g_{KK\omega}$, and $g_{\pi\pi\phi} = 0$.

    \item The use of G-Parity naturally produces the correct sign related to the anki-kaon potential depth and dispersion relation. Moreover, it gives us a clear picture of the role of each meson in the anti-kaon-anti-kaon interaction, as well as in the anti-kaon nucleon interaction.

    \item Different values of $\alpha_V$ produce different values of the kaon potential depth, which can potentially constrain the values of $\alpha_V$.

    \item For $N\bar{K}$ matter, the results are only weakly dependent on $\alpha_V$, and the EOS becomes softer as we move away from the SU(6) symmetry. For $NY\bar{K}$ matter, there is a competition between hyperons and anti-kaons. Moving away from the SU(6) favors the onset of $\bar{K}$ while suppressing the hyperons. This can be explained by a combined effect of a larger value of an attractive $g_{\bar{K}\bar{K}\omega}$, as well as a larger value of the mixed term $g_{\Sigma\Lambda\rho}$.
    
    \item In the presence of anti-kaons, hyperonic EOSs become only slightly stiffer as we reduce the value of $\alpha_V$. The well-known effect of stiffening the EOS by reducing $\alpha_V$ is strongly compromised.

\end{itemize}

%%%%%%%%%%%%%%%%%%%%%%%%%%%%
\textbf{Acknowledgements:} L.L.L.  was partially supported by CNPq (Brazil)
under Grant No 305347/2024-1.

%%%%%%%%%%%%%%%%%%%%%%%%%
\bibliography{aref}
%%%%%%%%%%%%%%%%%%%%%%%%%%%%%

\end{document}